\documentclass[1p]{elsarticle}
\usepackage{amssymb, graphicx}
\usepackage{pbox}
\usepackage{natbib}
\usepackage{url}

\begin{document}

\begin{frontmatter}

\title{Mantid - Data Analysis and Visualization Package for Neutron Scattering and $\mu SR$ Experiments}

\author[isis,tessellaUK]{O.~Arnold}
\author[ornl]{J.~C.~Bilheux}
\author[ornl]{J.~M.~Borreguero}
\author[isis]{A.~Buts}
\author[ornl]{S.~I.~Campbell}
\author[isis,ill]{L.~Chapon}
\author[ornl]{M.~Doucet}
\author[isis,tessellaUK]{N.~Draper}
\author[ill]{R.~Ferraz Leal}
\author[isis,tessellaUK]{M.~A.~Gigg}
\author[ornl]{V.~E.~Lynch}
\author[isis]{A.~Markvardsen}
\author[uws,ornl]{D.~J.~Mikkelson}
\author[uws,ornl]{R.~L.~Mikkelson}
\author[nccsornl]{R.~Miller}
\author[isis]{K.~Palmen}
\author[isis]{P.~Parker}
\author[isis]{G.~Passos}
\author[isis]{T.~G.~Perring}
\author[ornl]{P.~F.~Peterson}
\author[ornl]{S.~Ren}
\author[ornl]{M.~A.~Reuter}
\author[ornl]{A.~T.~Savici}
\corref{cor1}
\let\thefootnote\relax\footnotetext{For general Mantid correspondence use \texttt{mantid-help@mantidproject.org}}
\ead{saviciat@ornl.gov}
\author[isis]{J.~W.~Taylor}
\author[ornl,tessellaUS]{R.~J.~Taylor}
\author[isis,tessellaUK]{R.~Tolchenov}
\author[ornl]{W.~Zhou}
\author[ornl]{J.~Zikovsky}

\address[isis]{ISIS Facility, Rutherford Appleton Laboratory, Chilton, Didcot, Oxfordshire, UK}
\address[tessellaUK]{Tessella Ltd., Abingdon, Oxfordshire, UK}
\address[ornl]{Neutron Data Analysis and Visualization, Oak Ridge National Laboratory, Oak~Ridge,~TN,~USA}

\address[ill]{Institut Laue-Langevin, Grenoble, France}
\address[uws]{University of Wisconsin-Stout, Menomonie, WI, USA}
\address[nccsornl]{Computing and Computational Science Directorate, Oak Ridge National Laboratory, Oak~Ridge,~TN,~USA}
\address[tessellaUS]{Tessella Inc., Newton, MA, USA}

\begin{abstract}
The Mantid framework is a software solution developed for the analysis and visualization of neutron scattering and muon spin measurements. The framework is jointly developed by software engineers and scientists at the ISIS Neutron and Muon Facility and the Oak Ridge National Laboratory.  
The objectives, functionality and novel design aspects of Mantid are described.
\end{abstract}

\end{frontmatter}

\section{Introduction}
\label{Introduction}

The use of large scale facilities by researchers in the fields of condensed matter, soft matter, and the life sciences is becoming ever more prevalent in the modern research landscape. Facilities, such as Spallation Neutron Source (SNS) and High Flux Isotope Reactor (HFIR) at Oak Ridge National Laboratory, and ISIS at Rutherford Appleton Laboratory, have ever increasing user demand, and produce ever increasing volumes of data. One of the most important barriers between experiment and publication is the complex, and time consuming, effort that individual researchers apply to data reduction and analysis. Data reduction is the transformation of a dataset collected from an instrument into a dataset in physical units. This transformation requires detailed knowledge of the instrument. 
The objective of the Manipulation and Analysis Toolkit for Instrument Data\cite{mantiddoi} (Mantid) framework is to bridge this gap with a common interface for data reduction and analysis that is seamless between the user experience at the time of the experiment and at their home institute when performing the final analysis and fitting of data.

The main goals for the project are:
\begin{itemize}
\item To provide a technique independent, neutron and muon specific framework  to reduce, visualise and perform scientific analysis of data
\item To ensure quality by following professional software development practices
\item To actively support multiple platforms (Linux, OS X, Windows)
\item The software, source and documentation will be freely distributable
\item The framework must be extensible by instrument scientists and users
\item Provision of comprehensive, well maintained documentation
\end{itemize}

This paper contains a general description of the project and the main components of the Mantid software. A status of the progress in achieving these objectives is presented in the conclusions section.  

\section{General Description of the Mantid Project}
The Mantid project\cite{mantiddoi} is a large international collaboration between the Science and Technology Facilities Council (STFC) (UK) and the Department of Energy (DOE) (US) to co-develop a high performance computing framework for analysis of: powder and single crystal neutron diffraction data, inelastic and quasi-elastic neutron scattering data, polarised neutron diffraction data, neutron reflectometry data, small angle neutron scattering data and $\mu SR$ data. 
It was started in 2007 at ISIS, and joined by SNS and HFIR in 2010. More recently, contributions have been made by Institut Laue-Langevin (ILL), Paul Scherrer Institute (PSI), Bragg Institute at Australian Nuclear Science and Technology Organisation (ANSTO), and European Spallation Source (ESS).

In the past, each instrument (or instruments groups at a given facility) would develop individual bespoke software routines for their own science areas\cite{DAVE, OpenGenie, LAMP, ISAW}. Over the life of a facility ($>$40 years), this leads to a vast, unmanageable, library of mission critical software routines. Such a model is prone to single point failures. As individual authors of software leave a facility they take with them the key knowledge of the software they developed. This often leads to refactoring of existing code as the facility attempts to get back control of its mission critical software. 

Mantid has been developed with the overall objective of giving facilities and their users access to state of the art bespoke software that is professionally developed and maintained, with a clear science led strategic development and maintenance plan.
This methodology allows instrument scientists time to determine key software requirements for their user programs rather than having to develop and maintain software packages, in so doing both the user community and the facility benefit.

The overall ethos of the project is that of abstraction. That is to say, code developed within the project should at all times operate on all data types from all participating facilities. This idea leads to a framework that is, in principle, easier to use and maintain.

The Mantid framework consists of a highly modular C++/Python architecture which supports user built plug-in functions as well as access to powerful visualization toolkits such as ParaView\cite{paraview}. This modular design allows users to extend the capability of the framework to almost any application. The framework is provided under the GNU General Public Licence version 3\cite{gpl}, and is built for all commonly used operating systems.

\section{Neutron Scattering}

Neutron scattering is an established technique for determining the structure and dynamics of materials. It has generated a large user community, with research interests from life sciences to quantum magnetism. To meet the current and future demands in these areas, there have been a number of new large scale facilities built, or in the process of being built in the last 10 years. These new facilities are all pulsed spallation neutron sources rather than reactors. Pulsed spallation sources by definition have a time structure to the neutron production and, as a result of this, all instruments operate in a detection mode known as time of flight (TOF). TOF neutron instruments have the advantage of being able to collect data over a wide range in $S(\textbf{q},\omega)$ in a single pulse. 
In a neutron experiment one must relate measured counts to the physically meaningful $S(\textbf{q},\omega)$.

At a modern TOF neutron source it is common for instruments to have $10^5 n\ cm^{-1}s^{-1}$ and millions of pixels, generating GB size data files. In many experiments it is possible for several files to be combined together to create a large $n$ dimensional dataset or volume with a size of up to 1 TB.
Recently, pulsed sources have started to collect data in what is called event mode. This method simply lists to a file every detected neutron with a time of collection and other metadata. From the event list, one may filter based on time or metadata to create data subsets. This method has several advantages, it is effective for storing sparse data, and it allows time resolved experiments to be performed. 
Large data volumes, $n$ dimensional data and event mode format add several layers of complexity to the data reduction chain. For the instruments to be fully exploited, high performance software is a necessity.

\section{Muon Spin Relaxation/Rotation/Resonance ($\mu SR$) }

Muons provide a local probe to investigate the properties of a wide range of materials. 
$\mu SR$ has wide applicability and provides useful dynamic information for a broad range of science  from soft matter to quantum magnetism,  which is often complementary  to that from neutron scattering. 
The technique is similar to that of nuclear magnetic resonance, in which the polarisation of the target nuclei, in this case the muon, is tracked as a function of time. 
In the case of muons, spin polarised muons are implanted into the material under investigation and these muons decay into  positrons which are emitted preferentially along the final spin direction of the muon. 
By time stamping the detected positrons, the muon polarisation is inferred. As muons are produced by the decay product of pions, which in turn are produced by high energy protons ($\sim$800 MeV at ISIS), experiments are conducted at proton accelerators and are often situated next to spallation neutron sources, e.g. ISIS, PSI and J-PARC. 
This means that the users of neutron instruments often use muons as well and having a familiar software framework for analysis is clearly beneficial.
The Mantid framework fulfils this requirement, comprising  a wide range of methods with which to analyse the muon depolarisation spectrum:  integrated asymmetry, Fourier transform, maximum entropy and time domain analysis among others. 
Moreover, simulations of muon data using Density Functional Theory or electronic calculations can yield further insights into the material under investigation. 
The ability to link these simulations with the data analysis with a simple interface yields a very powerful tool for the analysis of muon experiments. Again, the Mantid framework offers this functionality to the instrument user.

\section{Development Practices}
\label{DevelopmentPractices}

One of the key  aspects of Mantid is the manner in which it is developed. 
In order to achieve the stated goals, a large team of about thirty scientists and scientific software engineers in Europe and United States are collaborating on this project. For an effective collaboration, we use several software development tools and practices designed to support distributed development teams. New feature requests or defect reports are entered into an issue tracking system. 

Another vital tool for organising work is the use of a version control system. Mantid uses git \cite{git} repositories for the source code, configuration files, and much of the documentation. To allow multiple developers to work in similar areas without interference, developers work on separate branches for each feature. To verify that there are no cross-platform compatibility issues, each feature branch is merged onto a `develop' branch whenever new code is ready. It is only after a feature branch has been completely addressed and tested that the code changes are merged onto the `master' branch from which release builds and new features are based.

In order to ensure quality, the Mantid project uses continuous integration. 
Whenever new code is committed to the `develop' branch, builds for each supported operating system are started, and are tested against a suite of over 6000 automated unit tests. A build is marked successful only if all of these unit tests pass.  
Once a day, a series of over 150 integration `system tests' are run with the most recent locally installed version. Builds that pass all system tests are immediately available for download and, in some cases, automatically deployed to computers. Formal releases of Mantid occur approximately every three months, and undergo additional manual testing. These releases are accompanied by detailed release notes and training.

\section{Mantid Design}
One of the main design consideration for this project was the separation of data and algorithms. The ethos of the development is that algorithms should (where possible) operate on all data types without \textit{a\ priori} knowledge of the data or the experiment that generated it. In principle, this ideology makes the framework cleaner and easier to use. In many instances, scientists are not experts in neutron scattering, $\mu SR$, or the associated data analysis that is required. Successful software application written for scientists must take this into account at the design stage.

\begin{figure}[!ht]
\centerline{\includegraphics[width=0.75\textwidth]{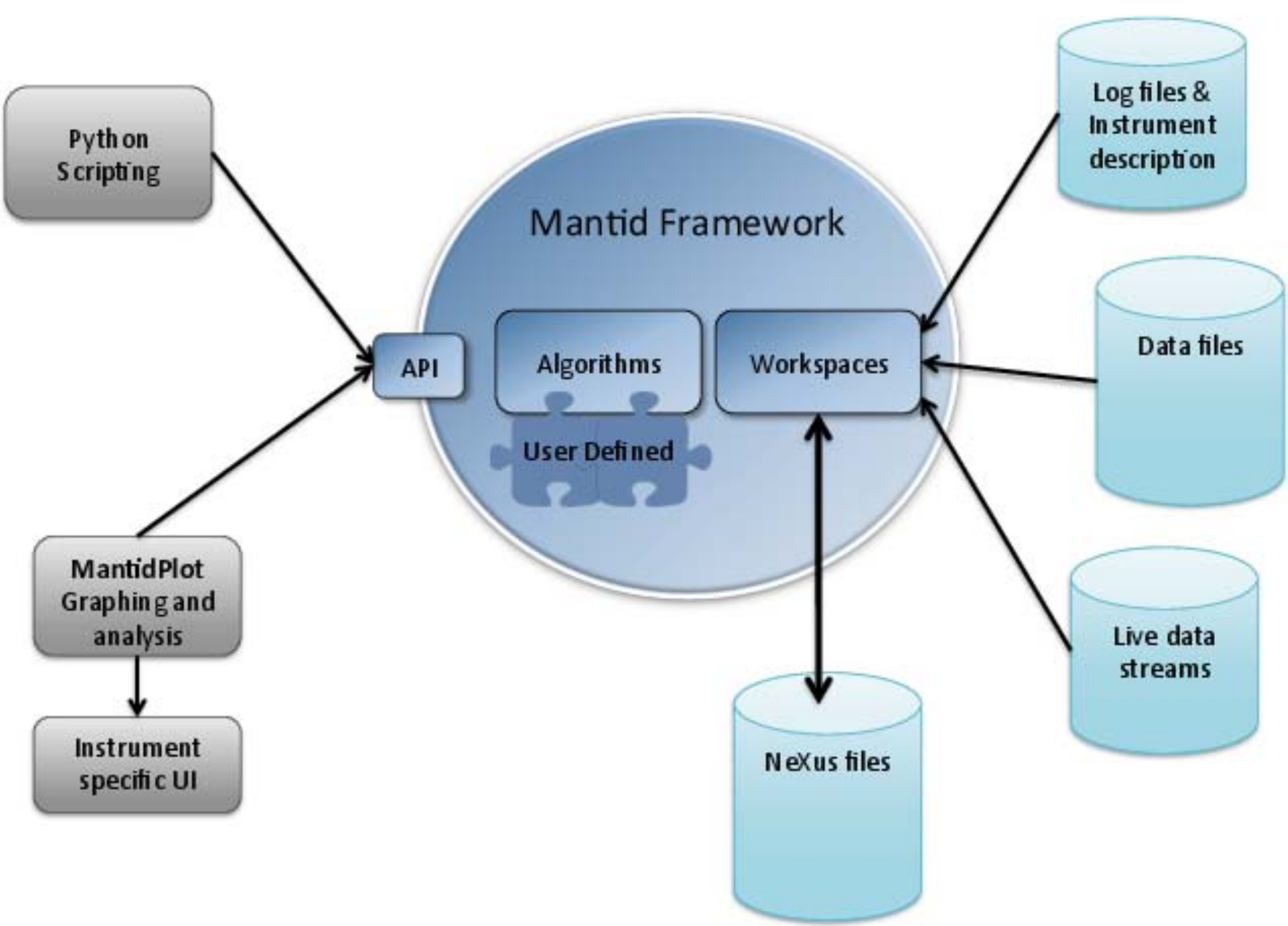}}
\caption{Mantid framework design}
\label{fig:Framework}
\end{figure}

Data containers (called workspaces) and algorithms, which manipulate workspaces, compose the central element of the Mantid framework (Figure \ref{fig:Framework}). 
Workspaces and algorithms are aware of the geometry of each individual instrument. 
Workspaces can be loaded from various file formats, live data streams, or created by different algorithms. The workspaces can be manipulated by the many algorithms in Mantid, and saved in a variety of formats. By default, Mantid uses the NeXus format\cite{NeXus} for saving intermediate and processed data, but various other output formats are also supported.

To ensure high performance for data analysis, but also allow flexibility in how the data is processed, the project is written in C++ with Python bindings. For parallel processing, Mantid uses OpenMP\cite{openmp}, Posix threads and MPI\cite{mpi}. The interaction with the Mantid framework occurs through the application programming interface (API). Currently the main interactions occur through either Python or through the graphical interface.

\subsection{Instrument Geometry}
A full description of the instrument is used within the Mantid framework. One way to specify the geometry is the instrument definition file (IDF). The IDF is a XML description of all pertinent instrument components. The instrument definitions conform to an XML schema (XSD). There are system tests that validate the IDF against the schema, and check if the instrument definitions can be loaded into Mantid. There is a python library that can be used to generate IDF files.
 
The IDF component description can be expanded upon, to increase the information level accessible to the Mantid framework. 
Previous applications for neutron scattering data analysis have generally only described instruments by their primary and secondary flight paths and detector angles. A full description, based on constructive solid geometry, allows complex visualization of the instrument and its detectors, along with the possibility to perform Monte Carlo simulations.
To account for moving instrument components, the instrument geometry is updated using log values. 

\subsection{Data Sources}
The Mantid framework is capable of reading from a variety of data sources. The most commonly used are data files written in the NeXus standard, which is instrument independent. However, the framework can read legacy files (e.g. ISIS raw files), as well as various ASCII formats. 
Alongside the standard loading of a pre-existing datafile, Mantid can also access the instrument data directly to provide real time display of detector counts and live `on the fly' data processing. 

\subsection{Workspaces}
Workspaces are the data containers in Mantid. In addition to the data, workspaces can hold other types of information, such as instrument geometry, sample environment logs, lattice parameters and orientation. Each workspace also holds a history of the algorithms that were used to create it. That way each workspace can show its provenance, and also regenerate the commands used to make it. Depending on the organization of the data, there are various types and subtypes of workspaces. More detailed information can be found in the online documentation \citep{webpage,doxygen}.

MatrixWorkspaces contain data for multiple spectra. A spectrum consists of an independent variable (e.g. time of flight, energy transfer), signal, and uncertainty. This is a common way to store histograms. 

\begin{figure}[!ht]
\centerline{\includegraphics[width=0.75\textwidth]{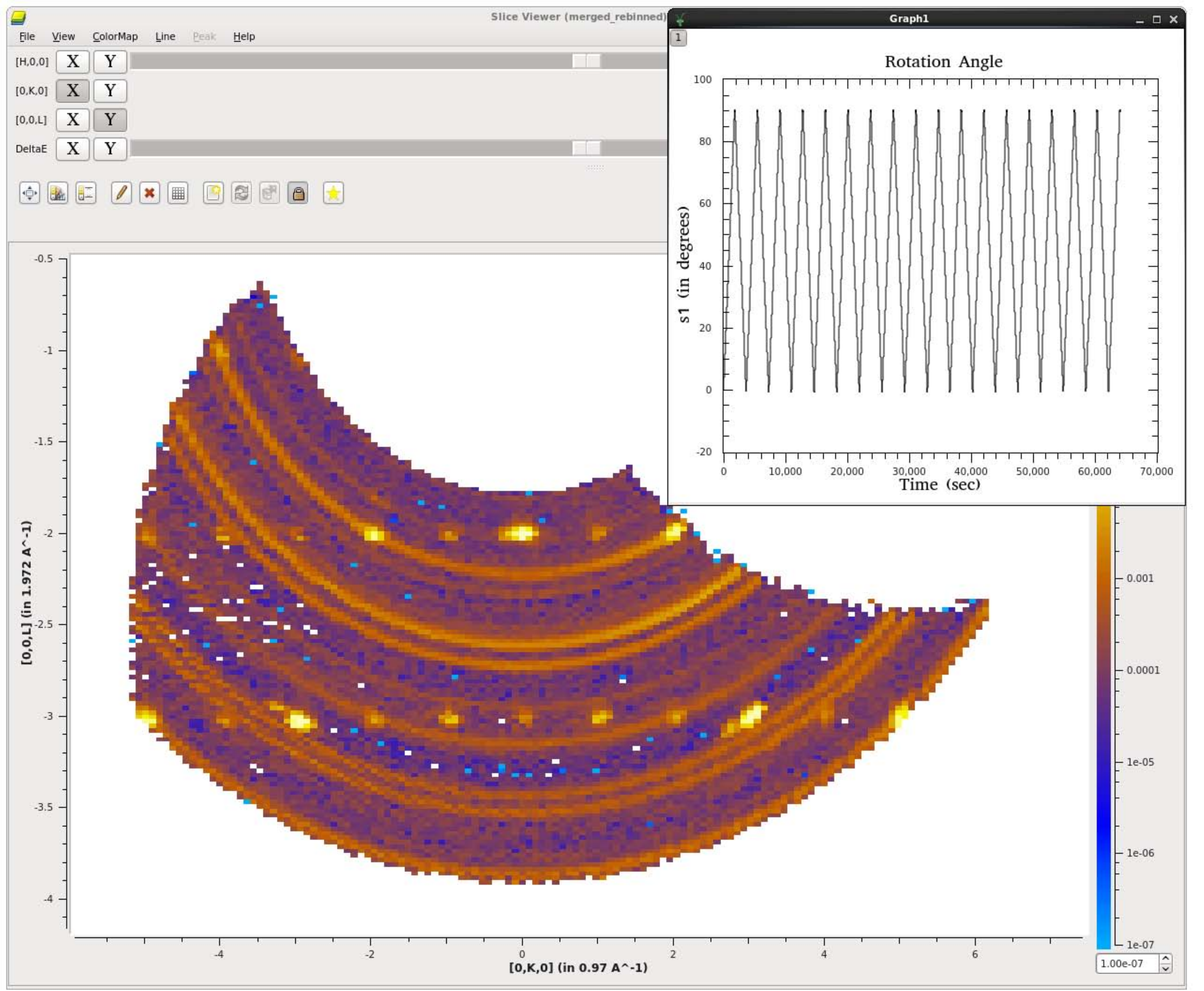}}
\caption{Continuous angle rotation example. Event data taken on HYSPEC spectrometer at SNS is filtered by angle, then converted to HKL momentum transfer components.}
\label{fig:ContinuousRotation}
\end{figure}

\begin{figure}[!ht]
\centerline{\includegraphics[width=0.75\textwidth]{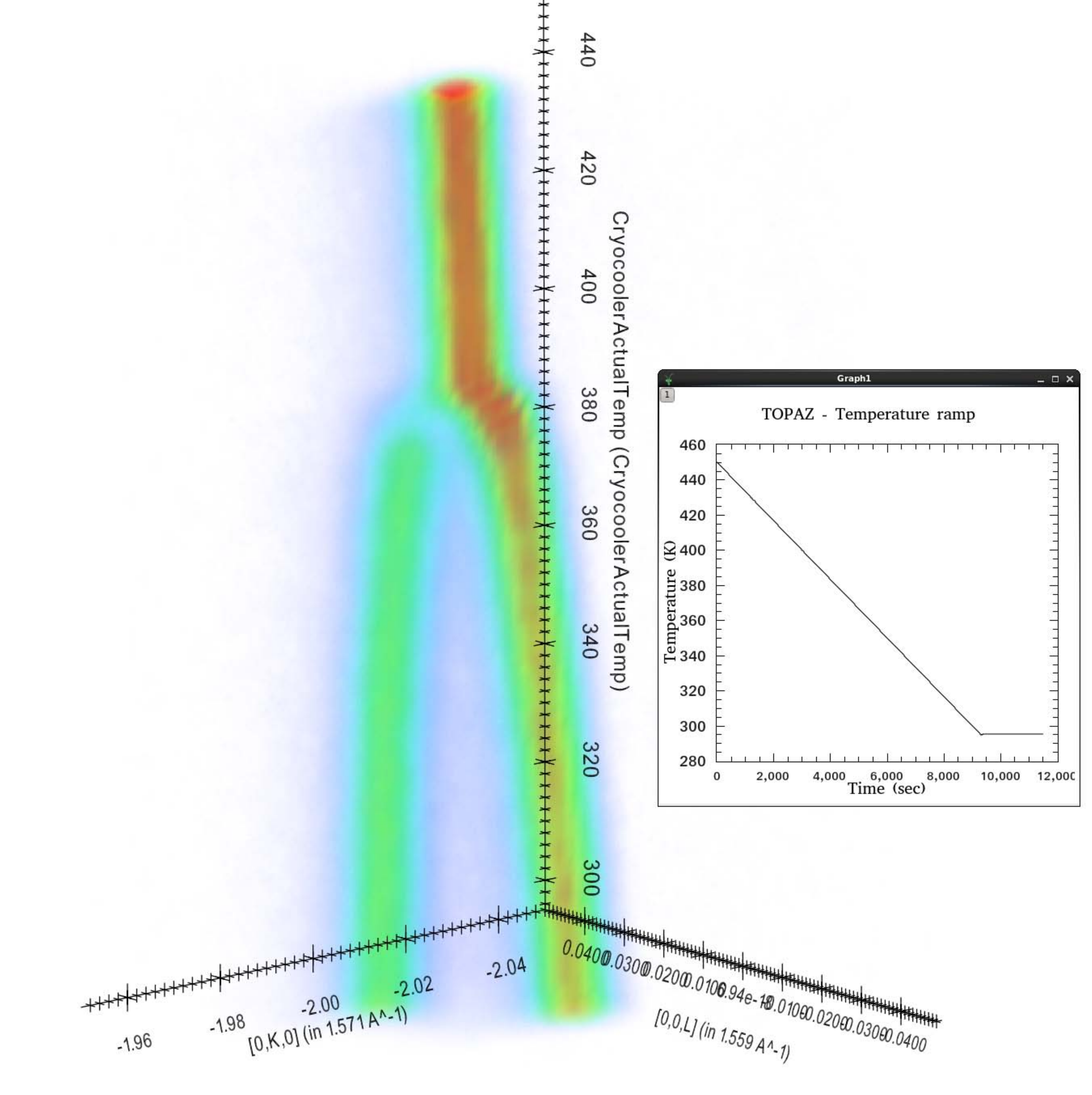}}
\caption{Example of event workspace usage for temperature ramp. Data taken on TOPAZ single crystal diffractometer at SNS shows a phase transition symmetry splitting on one of the Bragg peaks.}
\label{fig:TempRamp}
\end{figure}

The data acquisition system at several facilities now allow recording of each  detected neutron, labelling it with time-of-flight and wall-clock-time. In Mantid, this is stored as EventWorkspaces \cite{EventPaper}. 
EventWorkspaces also provide a histogram representation as well, which is calculated on the fly. This allows EventWorkspaces to be viewed as MatrixWorkspaces by the rest of the framework. The result is that algorithms and plotting work without the need to know the details of how data is stored. 
There are various uses for EventWorkspaces. One can filter out unwanted events, such as events recorded during temperature spikes. The other big use for events is allowing novel techniques, such as asynchronous parameter scans (continuous angle scans in Figure \ref{fig:ContinuousRotation}, temperature scans in Figure \ref{fig:TempRamp}), and pump probe experiments (pulse magnets, high frequency deformations of materials, and so on).

Another workspace type is the multi-dimensional workspace, or MDWorkspace. While for MatrixWorkspace there are two dimensions describing a data point (spectrum number and independent variable), for MDWorkspaces we have between 1 and 9 dimensions. Higher number of dimensions are required to accommodate labelling of data with extended parameter dependencies (e.g. sample environment variables). 

\begin{figure}[!ht]
\centerline{\includegraphics[width=0.75\textwidth]{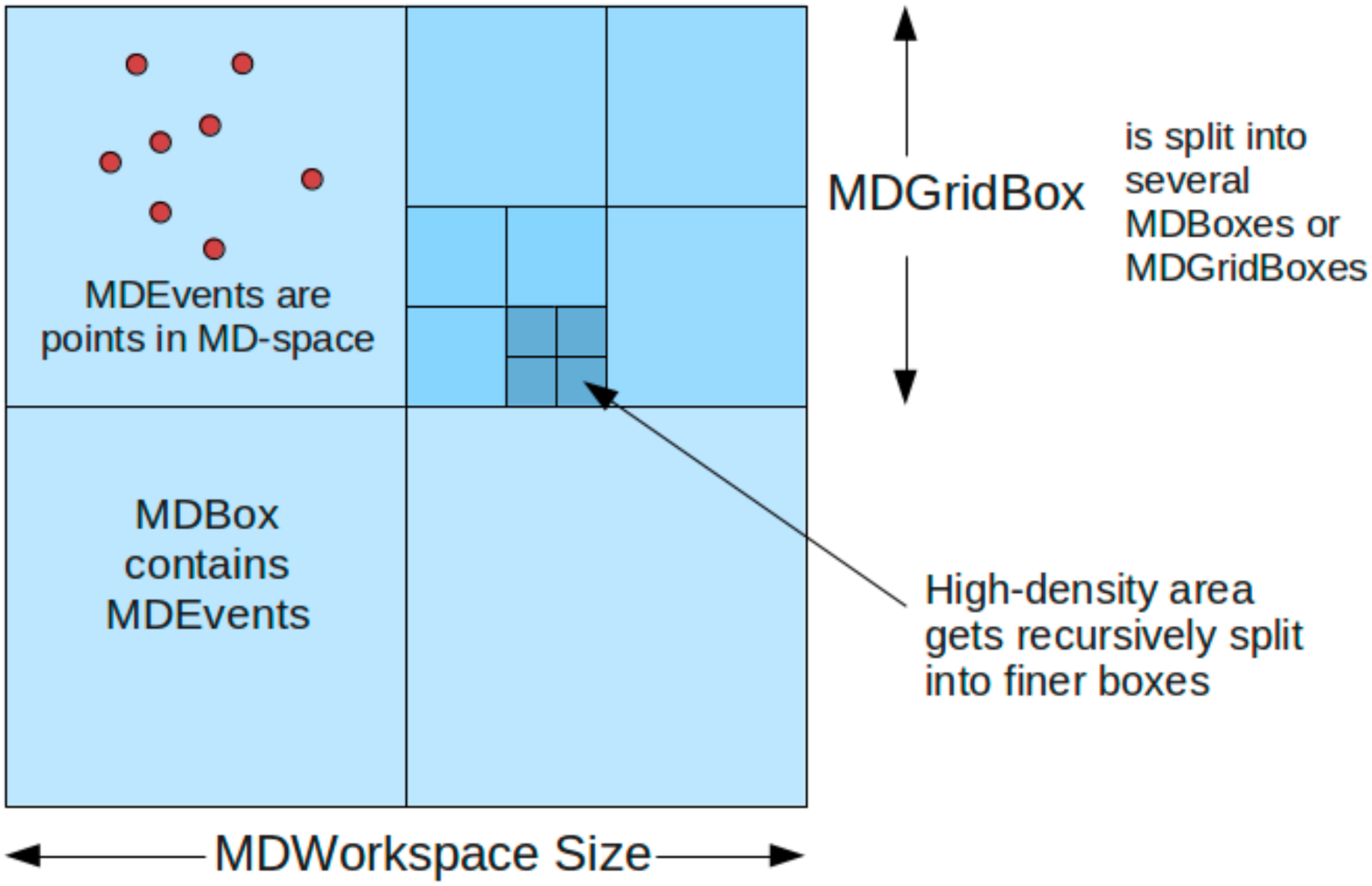}}
\caption{Schematic representation of the principle of adaptive rebinning used in the MDEventWorkspace type.}
\label{fig:MDEvent}
\end{figure}

For MDEventWorkspaces, each MDEvent contains coordinates, a weight and an uncertainty. It might also contain information about which detector and which run it comes from. All MDEvents are contained in MDBoxes. Above a certain threshold, the MDBox becomes an MDGridBox, by splitting into several equal size MDBoxes. This allows for an efficient searching and binning, and allows plotting on an adaptive mesh (see Fig. \ref{fig:MDEvent}).
MDHistoWorkspaces consist of signal and error arrays on a regular grid.

For data formats that contain different field types, Mantid provides TableWorkspaces. A TableWorkspace is organised in columns with each column having a name and type. Examples of TableWorkspaces are the parameters from model fitting, and a representation of information about Bragg peaks.

\subsection{Algorithms}
The algorithm layer is a key aspect of the Mantid framework.
Mantid algorithms are procedures to manipulate workspaces. They can be predefined, or written by users, in either C++ or Python.  The organization and development of algorithms is key to maintaining the ethos of the project. This presents a number of challenges for development as the framework can access multiple data types, from a variety of instruments. At the present, there are over 500 algorithms covering data handling (loading/saving workspaces from/to files), arithmetic operations (plus, minus, multiply, divide), unit conversions, and many technique specific algorithms (powder diffraction, single crystal diffraction, SANS, reflectometry, direct and indirect spectrometry, and $\mu SR$).  

The case of event mode data is interesting as it presents an efficient way of processing sparse data. It is often more efficient to keep the data as events through a chain of operations. This requirement has resulted in the development of a number of specialized event data handling operations. The end result is that for many reduction chains the data is events type until the final presentation. 

Core algorithms can be grouped together to form data reduction and analysis for individual instruments and science areas. These large algorithms can then be presented to the user at the Python scripting layer, command line interface or as a custom reduction user interface. 

In some cases a single `workflow algorithm' is beneficial, such as for live event process. The application can access the live data streams of event mode instruments at SNS and ISIS and can directly read histogram data from the detector electronics of ISIS instruments.

\subsection{Python API and Scripting}
The Python API provides an exceptionally powerful interface to Mantid. Many classes within the framework are open to Python control. The algorithms are added to the API at runtime, allowing new plugin algorithms to be available without further configuration. The Python API can be used to simply interact with existing functionality. Furthermore, Python can also be used to extend the capabilities of the Mantid framework by adding further algorithms or fit functions without needing to recompile or even restart the program.

The API has been written to give an intuitive Python feel, allowing a simple powerful syntax with minimal specific understanding of Mantid.  More advanced usage is possible within Python scripts, allowing popular packages such as NumPy, SciPy, matplotlib \cite{numpy,scipy, matplotlib} to be mixed with Mantid algorithms to process data.

\section{User Interface}
\subsection{MantidPlot}
\begin{figure}[!b]
\centerline{\includegraphics[width=0.75\textwidth]{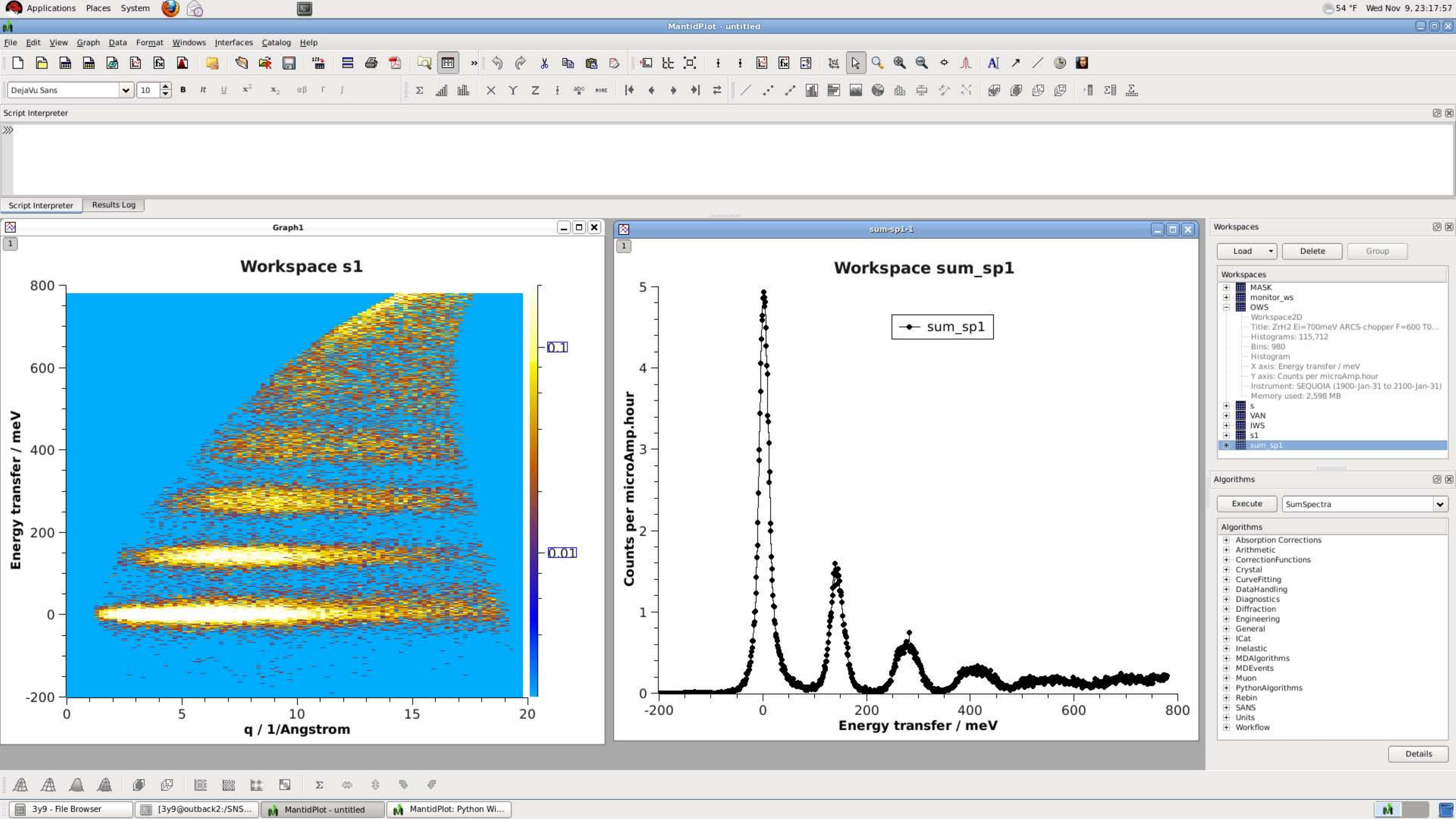}}
\caption{MantidPlot interface, showing 1D, and 2D plots. Lists of workspaces and algorithms are available on the right side. Data shown is from the SEQUOIA spectrometer at SNS.}
\label{fig:MantidPlot}
\end{figure}
The main interaction with Mantid occurs through the MantidPlot interface (Figure \ref{fig:MantidPlot}), based on QtiPlot\cite{qtiplot}. It allows visualizing and processing the data, Python scripting, and a generic fitting system.
A list of all algorithms, organized in categories, is also present. Clicking on an algorithm will open an automatically generated dialog box, with entries for each of the input parameters. A validation occurs when information is filled and any invalid input is flagged with an error message for the user. For each algorithm dialog box, a button allows for invoking the built-in help. A log window, where users can see the results of running different algorithms, is available. For several scientific techniques, custom interfaces are accessible from the MantidPlot menu.

The workspaces toolbox shows a list of all the workspaces currently available. Expanding the workspace entries show information about their type and content. A context sensitive menu allows plotting, instrument view, inspection of the sample environment logs, or the history of the workspace.

\subsection{Custom Interfaces}

Data reduction and basic analysis for individual instruments or science areas is generally a sequential chain of operations starting from data loading and resulting in a dataset that has meaningful units. As such, reduction for several scientific techniques can be complicated. More often than not, development of new features in this area must take into account legacy usage requirements and be well validated against existing "known" good results. In all cases, development of data reduction chains are tightly controlled and validated. 

One of Mantid's objectives is to provide scientists with a simple and efficient interface to allow them to analyse their data.  To achieve this for multiple science areas and instruments, a number of custom interfaces have been implemented. 
Science areas and instruments specifically supported by the Mantid framework can be seen in Table \ref{table:Coverage}.

\begin{table}[!htdp]

\footnotesize
\begin{tabular}{p{0.4\textwidth} |  p{0.5\textwidth} }
Science area &Instruments \\ \hline
Powder neutron diffraction&GEM HRPD WISH POLARIS POWGEN NOMAD VULCAN\\
Single crystal neutron diffraction&WISH SXD TOPAZ MANDI\\
Inelastic neutron scattering (direct)&MERLIN MAPS MARI LET SEQUOIA ARCS HYSPEC CNCS IN4 IN5 IN6 \\
Inelastic neutron scattering (indirect)&BASIS IRIS OSIRIS TOSCA VISION\\
Small angle neutron scattering &SANS2D LOQ EQ-SANS GP-SANS BIO-SANS D33\\
Neutron reflectometery&CRISP SURF POLREF INTER OFFSPEC REF\_L\\
$\mu SR$&MUSR HIFI EMU\\
\end{tabular}
\caption{Current science areas and instruments supported by the Mantid framework}
\label{table:Coverage}
\end{table}

\subsection {Fitting}

\begin{figure}[!ht]
\centerline{\includegraphics[width=0.75\textwidth]{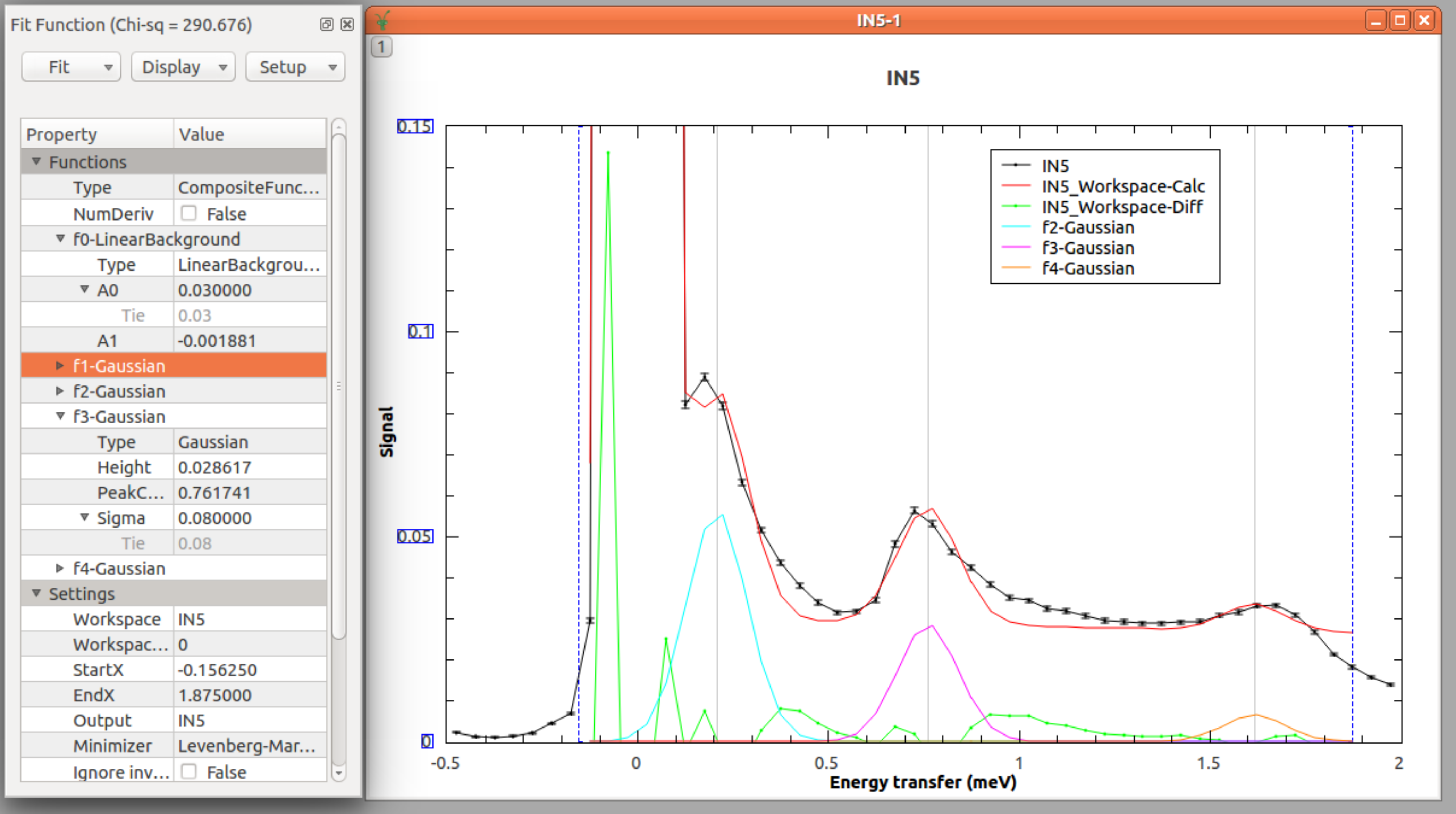}}
\caption{The simple fitting GUI interface in the MantidPlot application. Peak selection is performed using mouse selection on the displayed data.
The window on the left displays the required model and fit controls. The window on the right displays the data, fitted model, and difference. The vertical dotted lines indicate extents in x for the model. Data was measured on the IN5 spectrometer at ILL. }
\label{fig:Fitting}
\end{figure}
Fitting mathematical functions and models to experimental data is a key requirement of any scientific computing application. 
The Mantid framework has implemented a powerful engine for fitting multidimensional datasets. This is not intended to replace domain specific fitting software (e.g. GSAS, SASView). 
 Fitting peak functions and simple user derived functions to line data, i.e. data that is in 1D x,y,e format, can also be performed using a simple user interface (Figure \ref{fig:Fitting}). All of the fitting can be done within the Python scripting  interface.

Once the user has generated a model, the subsequent fitting can be batch processed across many different datasets, with the option of plotting fit results against a log parameter. Fitting results are displayed as TableWorkspaces which can then be further manipulated and analysed.

\subsection {Simulation and Analysis}

Fitting over multidimensional datasets is used in more complex situations, e.g. in the analysis of the results of the inelastic scattering experiments. Fitting a single resolution broadened model of scattering $S(\textbf{q},\omega)$ to a $n$ dimensional $S(\textbf{q},\omega)$ dataset is a standard data analysis procedure in this area used to account for substantial changes in the results of the experiment due to the instrument resolution effects.
 
Mantid contains a set of procedures for calculating an instrument resolution function and convoluting this resolution with chosen scattering model to obtain simulated resolution broadened scattering model. It then can use the multidimensional fitting framework to compare simulated model scattering with experimental scattering and fit the parameters of the scattering model to the results of the experiments. These capabilities are similar to the capabilities available in legacy programs (e.g. TobyFit \cite{tobyfit}, DAVE \cite{DAVE}). A Monte Carlo based instrument resolution model is implemented in Mantid. The framework allows defining and deploying other instrument resolution models. Mantid also contains a range of scattering models used in the analysis of the inelastic neutron scattering data.

\section{Visualization}

Modern instruments survey broad regions of reciprocal space, and therefore generate large data sets which cannot be easily visualized in 1D or 2D projections. 
Mantid provides a variety of tools for visualizing higher dimensional data.

\subsection{Instrument View}
\begin{figure}[!htb]
\centerline{\includegraphics[width=0.75\textwidth]{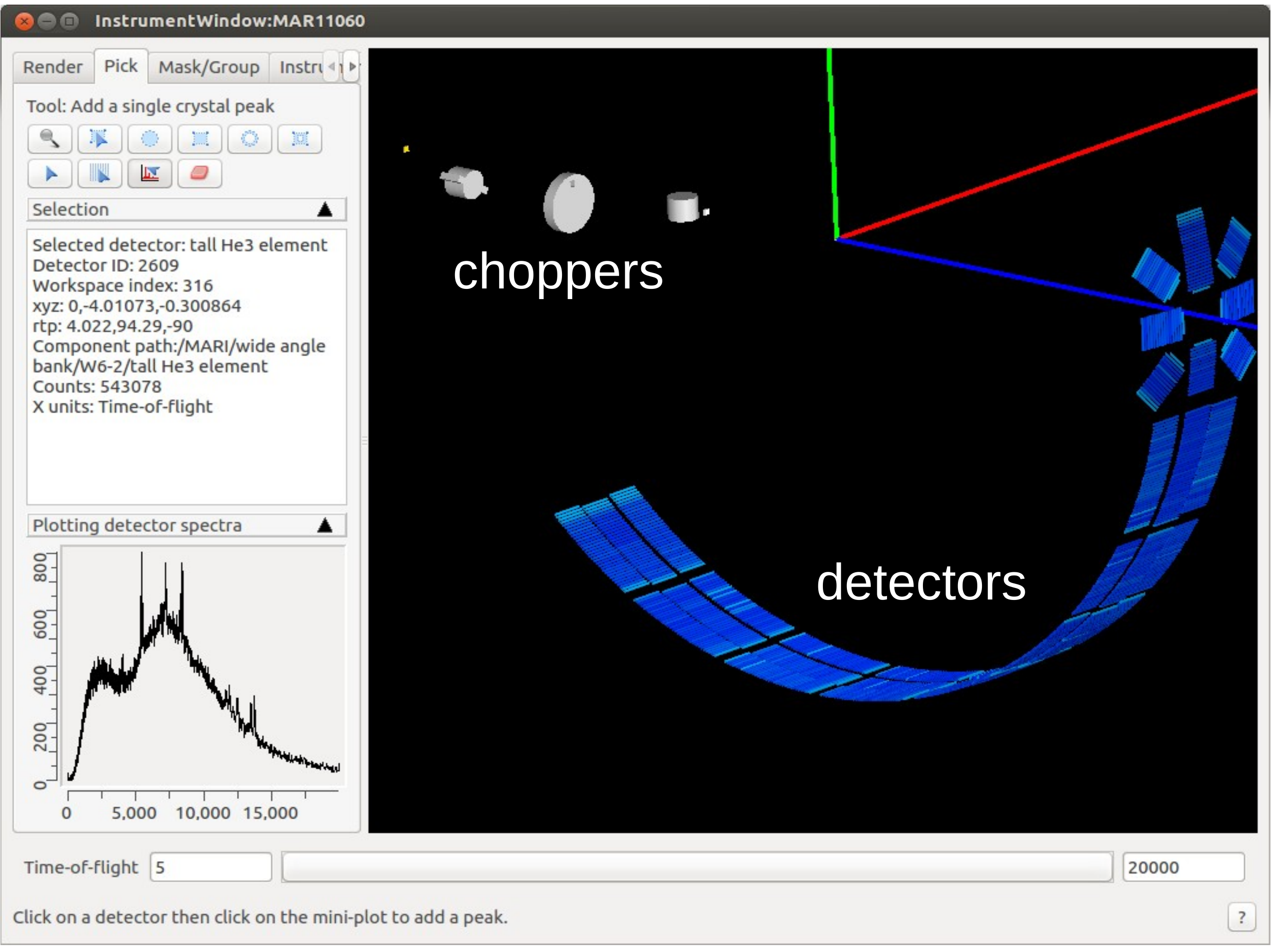}}
\caption{The Instrument View showing a 3D representation of the MARI spectrometer at ISIS. Various components are annotated.  }
\label{fig:InstrumentView}
\end{figure}

The Instrument View (Figure \ref{fig:InstrumentView}) is a 3D representation of the whole instrument, with component  positions  calculated from the IDF. Non-detector components (e.g. choppers, guides) can be toggled on or off. 
The colour of the detectors is representative of the total integrated counts. In addition to the 3D rendering, various 2D projections (e.g. spherical along $x$, cylindrical along $y$) of the detectors are available.
The  Instrument View allows for quick access to information about detectors, and provides a simple graphical interface for masking, grouping, and viewing spectra.

\subsection{Slice Viewer}

\begin{figure}[!htb]
\centerline{\includegraphics[width=0.75\textwidth]{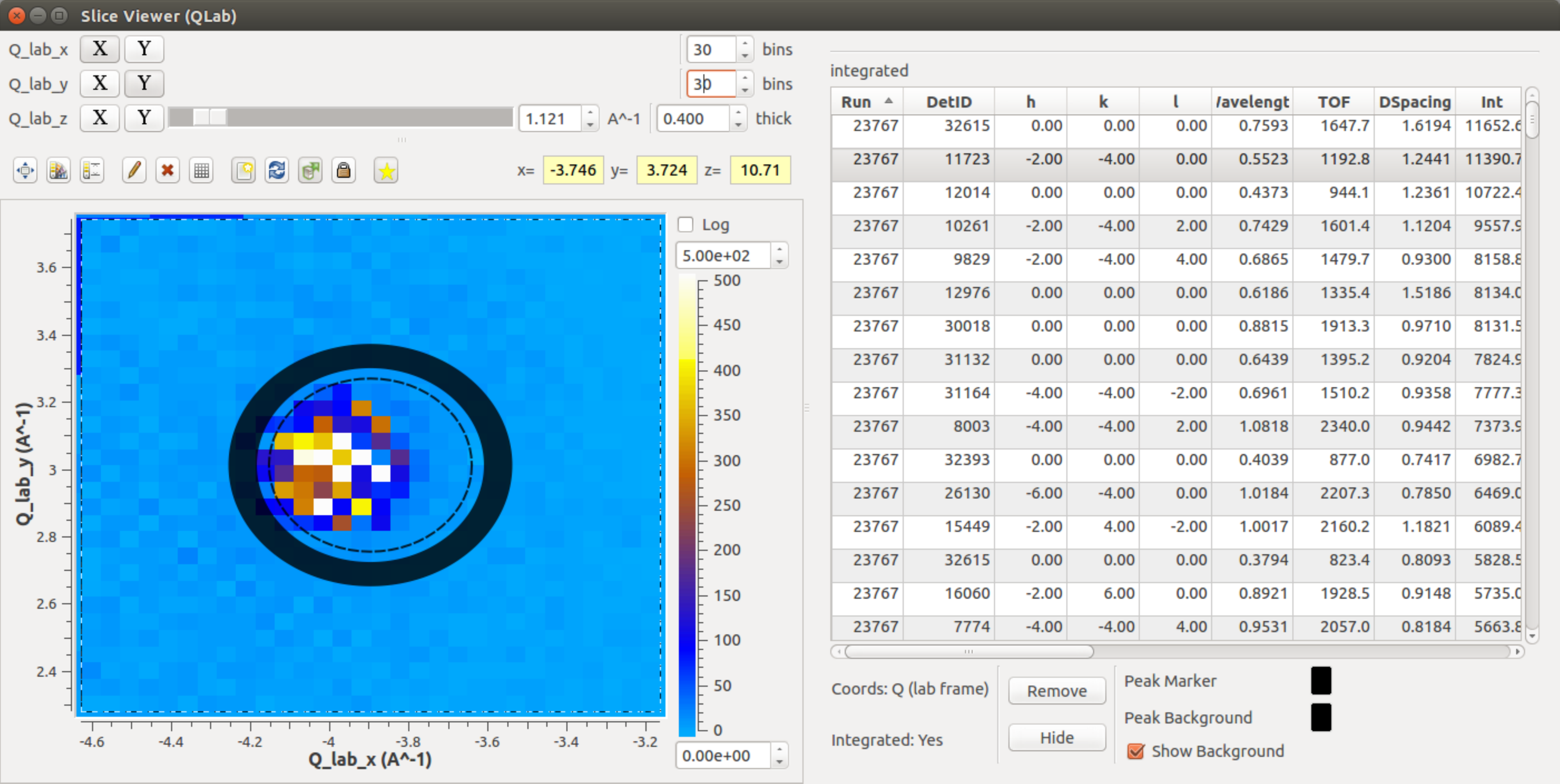}}
\caption{Slice Viewer showing a single crystal peak and related information. Data shown is NaCl, measured on SXD diffractometer at ISIS.}
\label{fig:SliceView}
\end{figure}

One tool for visualizing multidimensional (MD) data is the Slice Viewer (Figure \ref{fig:SliceView}). The Slice Viewer provides an interactive 2D projection of multiple data types.
Advanced features provide interactive line integration and overplotting PeaksWorkspaces. A list of overplotted peaks is available in this view.

\subsection{VATES Simple Interface}\label{VATES}

A major objective of Mantid has been the ability to represent multidimensional data \cite{DAVE, Mslice, Horace}. Originally the Visualization and Analysis Toolkit Extensions (VATES) project was an add-on to Mantid that is now fully integrated into the project. 
The VATES Simple Interface (\textit{VSI}), offers a limited set of data views and access to a subset of Mantid algorithms. It is based on application widgets and rendering libraries from the ParaView\cite{paraview} visualization program. The \textit{VSI}$\*$ takes advantage of the ParaView plugin architecture to provide functionality from within Mantid and from ParaView outside of Mantid. 

\begin{figure}[ht]
\centerline{\includegraphics[width=0.75\textwidth]{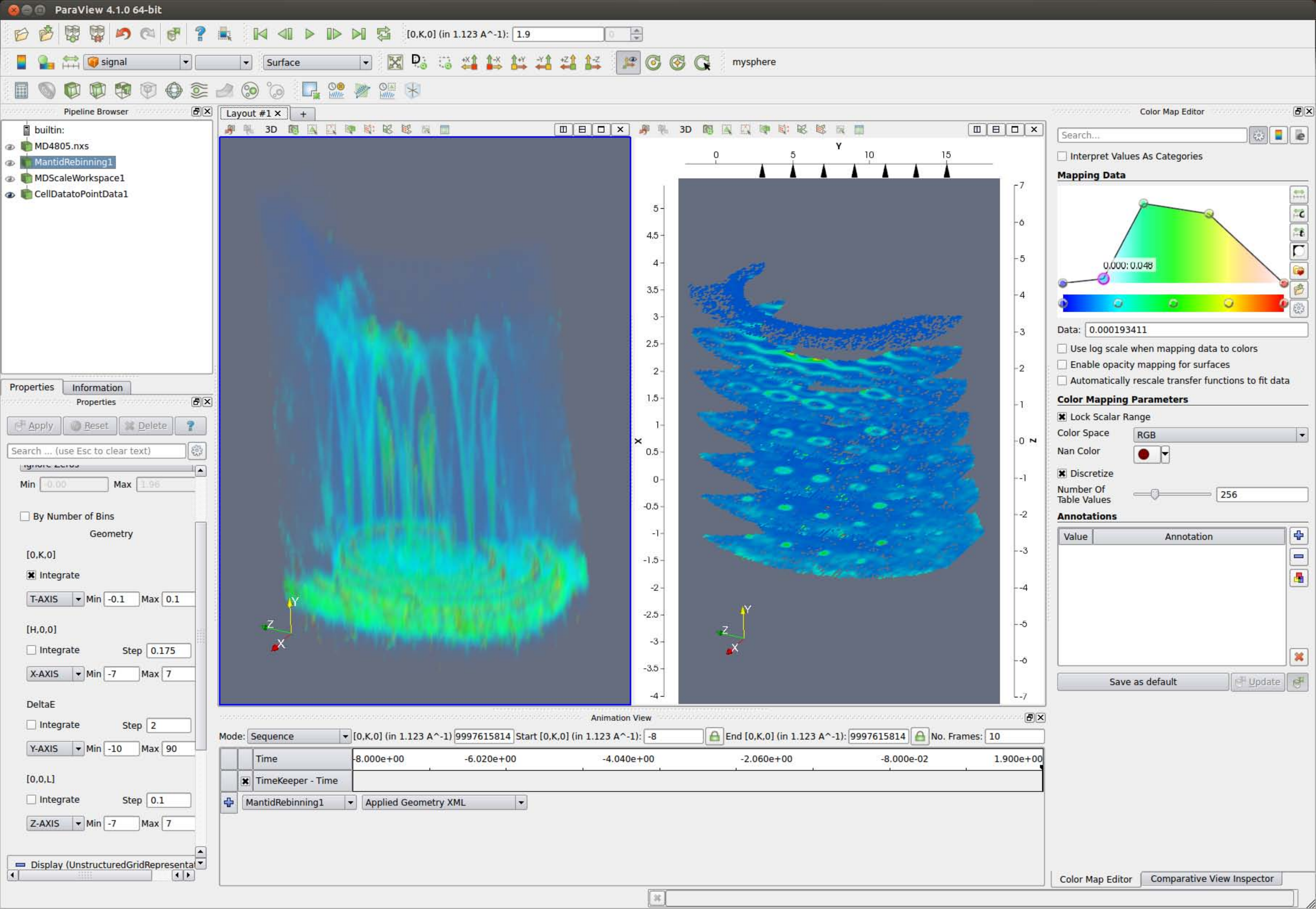}}
\caption{Paraview showing single crystal data\cite{YFeO3} on YFeO$_3$ from the SEQUOIA spectrometer at the SNS.}
\label{fig:PV}
\end{figure}

The data to be visualized passes through an API layer which translates the internal Mantid data structure to a VTK\cite{vtk} data structure, that can be rendered in the \textit{VSI}. Those same data structures can be saved to a file and visualized in the ParaView application. Indeed, it is possible to drive some aspects of multidimensional analysis directly from ParaView (Figure \ref{fig:PV} and Supplementary material (online only)\footnote{supplementary movies show inelastic neutron scattering in $YFeO_3$
\begin{itemize}
\item YFeO3\_slice.mp4: H, L, E volume at K=0, with a constant K=0, E slice at various energies
\item YFeO3\_varE.mp4: H, K, L volume at various energy transfer
\item YFeO3\_varK.mp4: H, L, E volume at various Ks
\end{itemize}
}. Data is from reference \cite{YFeO3}). The API layer provides the desired decoupling of the data structures and provides good flexibility to handle the various needs of the Mantid data structures and algorithms. 

\begin{figure}[!ht]
\centerline{\includegraphics[width=0.75\textwidth]{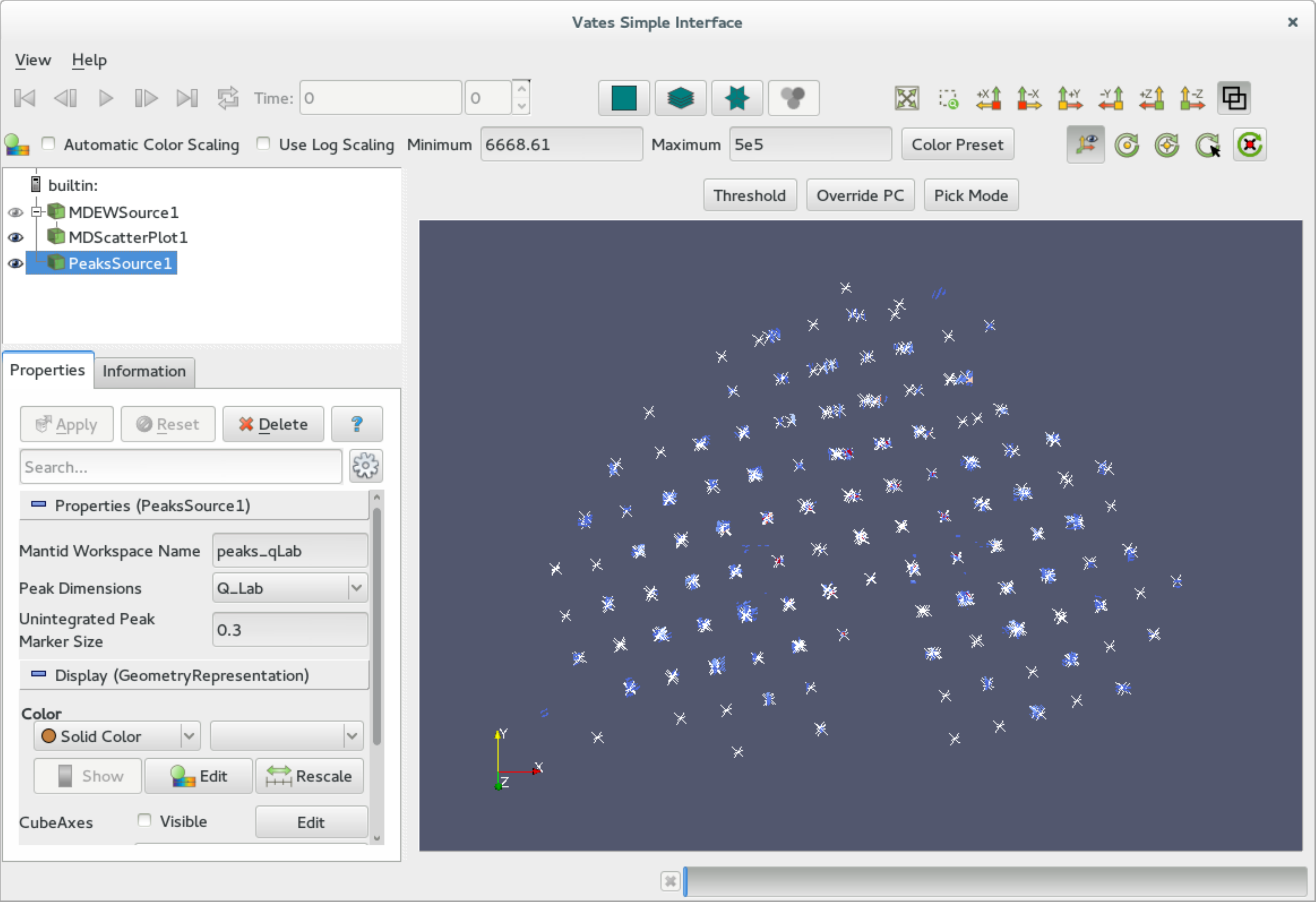}}
\caption{VSI in Splatter Plot mode with single crystal data from the SXD diffractometer at ISIS.}
\label{fig:VSI_sample}
\end{figure}

\begin{figure}[!ht]
\centerline{\includegraphics[width=0.75\textwidth]{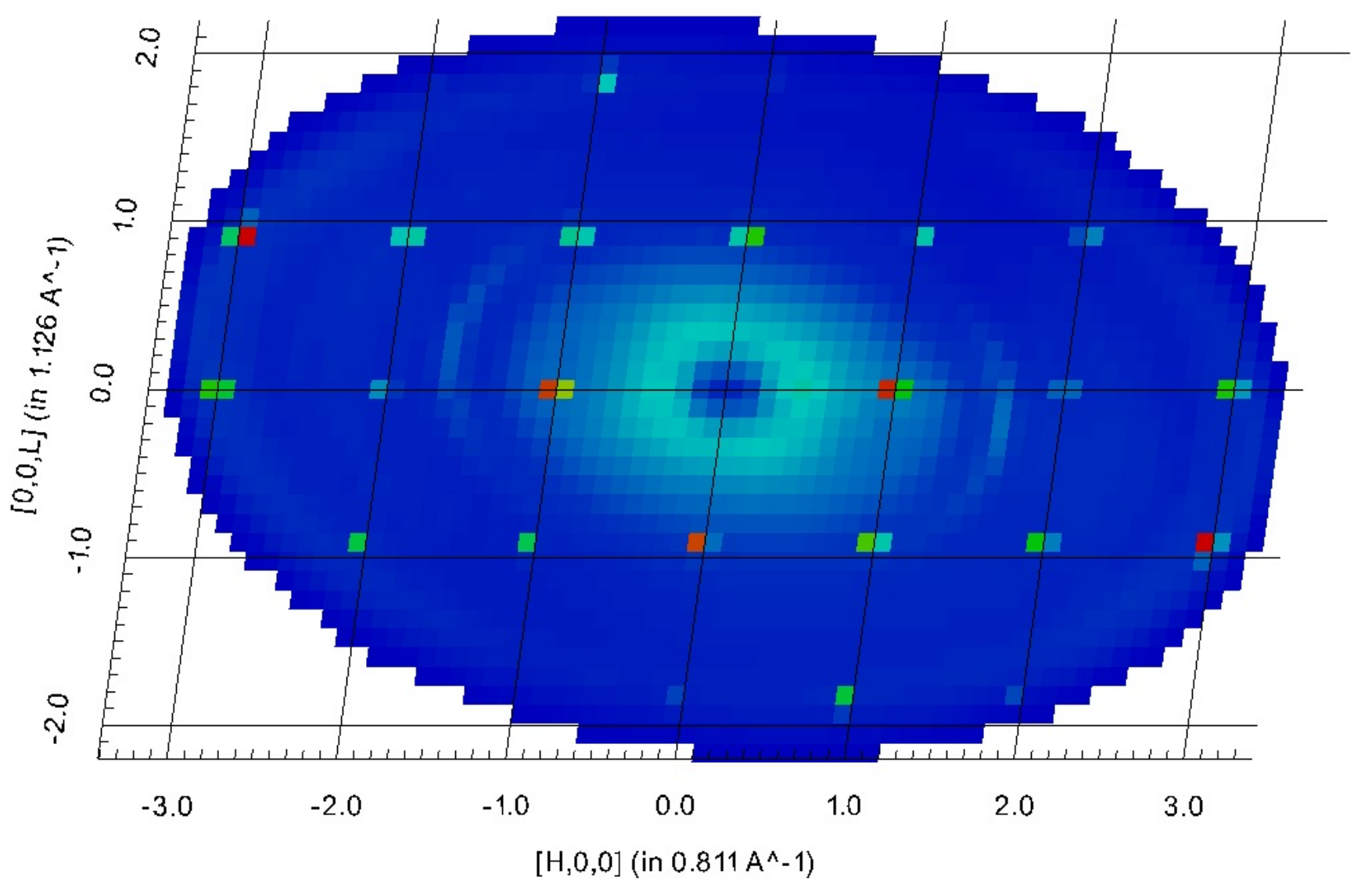}}
\caption{Diffraction pattern from a triclinic lattice, showing non-orthogonal axes. Data was measured on the CNCS spectrometer at the SNS.}
\label{fig:ParaView_sample}
\end{figure}

The \textit{VSI} has a view called Multi Slice which allows placing multiple orthogonal slices on the data. Those slices can then alternately be viewed in Slice Viewer for further exploration. The Splatter Plot (Figure \ref{fig:VSI_sample}) view is oriented towards visualizing peaks in single crystal diffraction data. In this view the user can interact with the data to retrieve information about a selected peak. The Three Slice view shows three orthogonal planes through the data with the capability exploring via moving a crosshair in one of the planes with a coordinate readout in each plane to show the location. The \textit{VSI}$\*$ has the ability to show the data with non-orthogonal axes such as the diffraction pattern for triclinic materials\cite{triclinic} in Figure \ref{fig:ParaView_sample}. This capability was implemented by Kitware\cite{kitware} via the SNS in support of the Mantid project.

\section{Community Involvement and Expandability}
The Mantid framework provides facility users with a very powerful data analysis tool. The Python API gives the user the ability to expand functionality for many different applications. User generated Python applications can be submitted to the Mantid script repository. The script repository  allows users to contribute and share scripts with the rest of the Mantid community, and MantidPlot allows uploading and downloading as well as marking scripts to be automatically updated with new versions from the  repository. The flexible nature of the framework can be used to analyse most types of experimental data, and is often used in new and interesting ways by the community that were not originally envisaged by the development team.
With the many algorithms supported by Mantid, extensive documentation is required.  This is provided at several levels, from helpful validation, intelligent code completion within the scripting environment, offline help provided with the installation and online help including examples and tutorials.
Finally the MantidPlot application allows users to submit bug reports, requests for assistance or just a suggestions for future development directly to the development team.

\section{Facility Integration}
\label{FacillityIntegration}
A very important step in Mantid development and deployment is facility integration.

To assist in this, Mantid interfaces with Information CATalog (ICAT) \cite{ICAT}. 
It is in use at both ISIS and SNS. Each facility uses a different approach to storing their archived data files. Mantid allows a small archive search adapter to be written to a provided interface, to locate raw or processed files in data archives at each of the facilities.

One important use of the ICAT interface is the autoreduction process on certain instruments. As soon as files are created and catalogued, a reduction script is automatically invoked. This script uses metadata in the file and/or the ICAT catalogue to reduce the raw data to a form that users are interested in. 

In addition, at SNS and ISIS, the development team has implemented an interface between Mantid and the data acquisition systems, in order to allow users to look and analyse their data in near real time. This approach allows for processing of the live data into scientifically useful results. This level of near real time data analysis allows for much more efficient use of valuable experiment time.

\section{Conclusion}

The Mantid project offers an extensible  framework (through Python) for data manipulation, analysis and visualization, geared toward neutron scattering and $\mu SR$ experiments. It is the main reduction software in use at SNS and ISIS, and partially in use or considered for widespread adoption at several other neutron facilities. Significant progress is being made on the analysis and visualization sides.
The ongoing goal of improving performance, usability, and documentation is helped by the development practices described in a section \ref{DevelopmentPractices}, and the community support and feedback outlined in section \ref{FacillityIntegration}.
The source code can be found at Github\cite{github}, and binary installers for Linux, OS X, and Windows can be found on the Mantid webpage\cite{webpage}.
Up to date information can be found in the offline help, and usage tutorials can be found on the Mantid web page\cite{webpage}. 

\section{Acknowledgements}
The development team would like to thank all instrument scientists and students at ISIS and SNS for their feedback and contributions, P. Radaelli, and R. McGreevy for championing the project at ISIS in the initial stages, and R. McGreevy, I. Anderson, and M. Hagen for forging the collaboration between ORNL and STFC. We acknowledge A. Hillier for contributions to this manuscript. Work at ORNL was sponsored by the Scientific User Facilities Division, Office of Basic Energy Sciences, US Department of Energy. Work at the ISIS facility was funded by the Science and Technology Facilities Council (STFC) UK. Development for ILL instruments was funded by NMI3 (WP6).

\begin{center}
\line(1,0){250}
\end{center}


\bibliography{Mantid}{}
\bibliographystyle{elsarticle-num}
\end{document}